\documentclass[%
 reprint,
 amsmath,amssymb,
 aps,
]{revtex4-2}

\usepackage{graphicx}
\usepackage{dcolumn}
\usepackage{bm}
\usepackage{hyperref}

\usepackage{xcolor}

\begin{document}

\title{Topological heavy-flavor tagging and intrinsic bottom at the Electron-Ion Collider}

\author{Thomas Boettcher}
 \email{boettcts@ucmail.uc.edu}
 \affiliation{Department of Physics, University of Cincinnati, Cincinnati, Ohio 45221, USA}

\date{February 16, 2024}

\begin{abstract}
Heavy-flavor hadron production, in particular bottom hadron production, is
difficult to study in deep-inelastic scattering (DIS) experiments due to small
production rates and branching fractions. To overcome these limitations, a
method for identifying heavy-flavor DIS events based on event topology is
proposed. Based on a heavy-flavor jet tagging strategy developed for the LHCb
experiment, this algorithm uses displaced vertices to identify decays of
heavy-flavor hadrons. The algorithm's performance at the Electron-Ion Collider
is demonstrated using simulation, and it is shown to provide discovery potential
for non-perturbative intrinsic bottom quarks in the proton.
\end{abstract}

\maketitle

\section{Introduction}
\label{sec:intro}

\begin{figure}
    {\centering
        \includegraphics[width=0.45\textwidth]{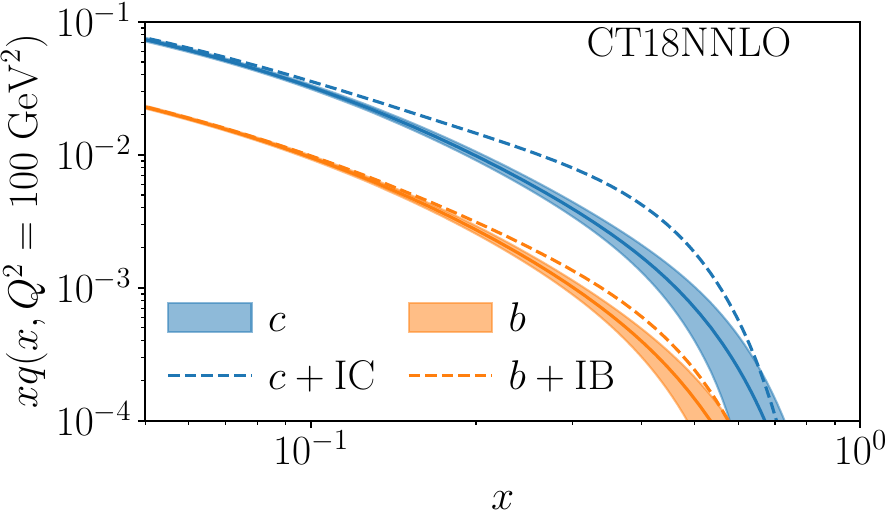}
        
    } \caption{Intrinsic charm and bottom PDFs. The baseline PDFs are from the
    CT18NNLO PDF set\,\cite{Hou:2019efy}. The shaded regions show the $68\%$
    confidence-level regions. The $c+{\rm IC}$ PDF is from
    CT18FC\,\cite{Guzzi:2022rca}. The $b+{\rm IB}$ PDF is obtained by scaling
    the intrinsic component of the CT18FC charm PDF by $m_c^2/m_b^2$ and adding
    the result to the baseline $b$ PDF.}
    \label{fig:ct18_ic_ib}
    
\end{figure}

The possible existence of non-perturbative ``intrinsic'' heavy quarks in the
proton was first proposed shortly after the discovery of heavy quarks
themselves\,\cite{Brodsky:1980pb}. %
Intrinsic heavy quarks are predicted to arise from a
$\left|uudQ\overline{Q}\right>$ component of the proton's wavefunction, where
$Q\overline{Q}$ denotes a heavy quark-antiquark pair. %
Various models predict the intrinsic contribution to the heavy-quark parton
distribution functions (PDFs), including models inspired by light-front quantum
chromodynamics (LFQCD)\,\cite{Brodsky:1980pb} and fluctuations of the proton into
heavy meson-baryon pairs\,\cite{Hobbs:2013bia}. %
These models generally agree that intrinsic heavy quarks carry a large fraction
$x$ of the proton's momentum, resulting in valence-like heavy quarks. %
This can be seen in Fig.~\ref{fig:ct18_ic_ib}, which shows LFQCD-inspired models
of intrinsic charm (IC) and intrinsic bottom (IB)\,\cite{Guzzi:2022rca}. %

Experimental searches for IC have been carried out in both fixed-target
deep-inelastic scattering (DIS) and high-energy hadron collisions. %
Charm structure function data from the European Muon Collaboration (EMC)
experiment\,\cite{EuropeanMuon:1982xfn} and studies of $Z$-boson production in
association with charm-quark jets (\mbox{$Z+c$}) by the LHCb
experiment\,\cite{LHCb:2021stx,Boettcher:2015sqn} are expected to be
particularly sensitive probes of IC. %
The LHCb experiment has also searched for evidence of IC in charm production and
charge asymmetry measurements in fixed-target proton-nucleus
collisions~\cite{LHCb:2018jry,LHCb:2022cul}. Intrinsic heavy flavor is typically
characterized by the average momentum carried by the intrinsic heavy quarks,
$\left<x\right>_{\rm IC,IB}$, at an initial energy scale \mbox{$Q_0=m_c$}. %
The NNPDF collaboration performed a global analysis including EMC and LHCb
\mbox{$Z+c$} data\,\cite{Ball:2022qks}. The analysis claimed $3\sigma$ evidence
for nonzero IC with $\left<x\right>_{\rm IC}\approx1\%$. %
A global analysis based on the CT18 PDF fit omitted the the LHCb and EMC
measurements due to difficulties with theoretical interpretation. %
The resulting fit mildly prefers nonzero IC, with $\left<x\right>_{\rm
IC}\approx0.5\%$\cite{Guzzi:2022rca}. %
Yet another global analysis excluded percent-level IC at the $4\sigma$
level\,\cite{Hobbs:2013bia}. %
The Electron-Ion Collider (EIC), under construction at Brookhaven National
Laboratory, is expected to produce in excess of 100 times more data than
previous collider DIS facilities, allowing for detailed studies of the charm
quark PDF\,\cite{AbdulKhalek:2021gbh}. %
Recent studies indicate that the EIC will be able to conclusively observe or
exclude percent-level IC in the proton\,\cite{NNPDF:2023tyk,Kelsey:2021gpk}. %

In contrast to the experimental and theoretical interest in IC, the possibility
of intrinsic bottom quarks in the proton has received relatively little
attention (see Ref.~\cite{Brodsky:2015fna} for a review). %
The size of the intrinsic heavy-quark contribution to the proton PDF is expected
to scale as $1/m_Q^2$, where $m_Q$ is the heavy quark mass, suppressing IB by an
order of magnitude relative to IC\,\cite{Lyonnet:2015dca}. %
As a result, both the absolute size of the IB contribution and its size relative
to the perturbative $b$-quark PDF are smaller than the analagous IC
contributions. %
The $b$-hadron cross section in DIS is also suppressed relative to the
$c$-hadron cross section due to the smaller electric charge of the $b$ quark. %
Additionally, the largest $b$-hadron branching fractions to fully
reconstructible final states are $\mathcal{O}(10^{-3})$\,\cite{ParticleDataGroup:2022pth}. %
As a result of these limitations, little data constraining the $b$-quark PDF exists. %
What little data does exist does not probe the valence region, leaving the IB
content of the proton almost entirely unconstrained\,\cite{H1:2018flt}. %
Consequently, no global analysis of IB in the proton has been performed. %

The experimental challenges of studying $b$-hadron production in DIS can be
partially overcome by using the topology of heavy-flavor hadron decays. %
This strategy was used by both the H1 and ZEUS experiments at HERA, which used
displaced tracks and secondary vertices to identify $b$-hadrons and extract the
$b\overline{b}$ contribution to the proton structure function,
$F_2^{b\overline{b}}$\,\cite{H1:2018flt,H1:2009uwa,ZEUS:2014wft}. %
The LHC experiments use a similar strategy to identify heavy-flavor jets. Jets
containing heavy-flavor hadrons are identified using the properties of displaced
charged-particle
vertices\,\cite{LHCb:2015tna,LHCb:2021dlw,ATLAS:2022qxm,CMS:2017wtu}. %
Using this strategy, the LHCb experiment is able to identify or ``tag'' about
$60\%$ of jets containing $b$-hadrons and distinguish between $b$ and $c$ jets.
The proposed detector at the EIC is expected to have vertex reconstruction
capabilities similar to those of LHCb, enabling a similar strategy for tagging
heavy-flavor DIS events\,\cite{AbdulKhalek:2021gbh}. %
Previous studies have explored the performance of topological charm tagging at
the EIC, but studies involving $b$ hadrons have focused on using fully
reconstructed decays to study hadronization\,\cite{Wong:2020xtc}. %
Previously explored charm tagging methods rely on the ability to identify
charged kaons or count displaced tracks, either in the entire event or clustered
into jets\,\cite{Arratia:2020azl,Dong:2022xbd,Aschenauer:2017oxs}. %
In contrast, the algorithm employed by LHCb does not require particle
identification and relies only on the topological properties of charged particle
vertices. %

This paper demonstrates how the LHCb experiment's jet tagging strategy can be
applied to study heavy-flavor production at the EIC. %
Because the LHCb jet-tagging algorithm depends only on the properties of the
heavy flavor decay and not on the jet itself, the algorithm can be naturally
adapted to identifying heavy-flavor events in DIS. %
Section~\ref{sec:sim} describes the simulation setup used for these studies, and
Section~\ref{sec:tag} describes the heavy-flavor tagging algorithm. %
Section~\ref{sec:ib} presents the expected sensitivity to IB, and
Section~\ref{sec:conc} summarizes conclusions and discusses additional uses for
topological heavy-flavor tagging at the EIC. %

\section{Simulation}
\label{sec:sim}

The tagging algorithm performance studies were conducted using simulated
\mbox{$e+p$} DIS events generated using the PYTHIA 8.3
generator\,\cite{Bierlich:2022pfr}. %
The simulation includes both neutral- and charged-current DIS, although the
charged-current contribution to the simulated samples is negligible. %
Simulations were performed for four beam energy configurations:
\mbox{$5\times100$}, \mbox{$10\times100$}, \mbox{$10\times275$}, and
\mbox{$18\times275~{\rm GeV}$}\footnote{Natural units are used throughout this
paper.}, where the first number of each pair is the electron energy and the
second is the proton energy.
These configurations correspond to \mbox{$\sqrt{s}=45$}, $63$, $105$, and
$141~{\rm GeV}$, respectively. %
Heavy-flavor events are defined by the presence of a heavy-flavor hadron. %
A $b$ event contains a $b$ hadron, whereas a $c$ event contains a $c$ hadron and
no $b$ hadron. %
A light-parton ($uds$) event contains no $c$ or $b$ hadrons. %

The tagging algorithm's performance was studied as a function of the kinematic
variables $x$ and $Q^2$. %
These variables can be used to calculate the inelasticity \mbox{$y=Q^2/(xs)$}.
The accessible kinematic region of interest for IB is \mbox{$Q^2>100~{\rm
GeV}^2$} and \mbox{$x>0.1$}. %
For the beam configurations used in this study, this kinematic region
corresponds to \mbox{$0.01\lesssim y\lesssim0.5$}, a region where the EIC detector is
expected to determine $x$ and $Q^2$ from the scattered electron with high
precision\,\cite{Armesto:2023hnw}. %
As a result, $x$ and $Q^2$ were determined at parton level for this study. %
Furthermore, radiative corrections are expected to be less significant for
heavy-quark production than for inclusive DIS and were ignored in this
study\,\cite{Aschenauer:2017oxs}. %

The response of a hypothetical EIC detector is modeled according to
parameterizations based on the expected performance of the future
detector~\cite{AbdulKhalek:2021gbh}. %
The momentum and position resolutions are given as functions of transverse
momentum ($p_{\rm T}$) and pseudorapidity ($\eta$), as shown in
Table~\ref{tab:smear}. %
Only long-lived charged particles with \mbox{$p_{\rm T}>200~{\rm MeV}$} and
\mbox{$|\eta|<2.5$} were considered for this study. %
A charged particle reconstruction efficiency of $90\%$ was assumed for the
entire fiducial region. %

The position of the collision vertex, or primary vertex (PV), is determined by
smearing the true position of the interaction point. %
The PV resolution is shown in Ref.~\cite{Kelsey:2021gpk} and estimated here as
\mbox{$\sigma_{x,y,z}=(10\oplus30/\sqrt{n})$}, where $n$ is the number of reconstructed
prompt charged particles. %
Reconstructed particles are classified as prompt based on
\begin{equation}
\chi^2_{\rm DCA,IP}=\frac{d_x^2}{\sigma_x^2}+\frac{d_y^2}{\sigma_y^2},
\label{eqn:dcachi2}
\end{equation}
where $d_{x,y}$ is the distance of closest approach of the smeared charged
particle to the interaction point in the dimension denoted by the subscript, and
$\sigma_{x,y}$ is the corresponding detector resolution determined using the
parameterizations from Table~\ref{tab:smear}. %
Tracks are considered prompt if \mbox{$\chi^2_{\rm DCA,IP}<12$}. %

\begin{table}

\caption{Resolution functions used to smear the generated charged particles to
simulate the EIC detector's response. Both $p$ and $p_{\rm T}$ are in GeV.}
\begin{tabular}{c c c c}
    \hline
    & $\sigma_{p}/p$ & $\sigma_{xy}$ & $\sigma_z$\\
    \hline
    $|\eta|<1$ & $0.04p\oplus1\%$ & $30/p_{\rm T}\oplus5~{\rm \mu m}$ & $30/p_{\rm T}\oplus5~{\rm \mu m}$\\
    $|\eta|<2.5$ & $0.04p\oplus2\%$ & $40/p_{\rm T}\oplus10~{\rm \mu m}$ & $100/p_{\rm T}\oplus20~{\rm \mu m}$\\
    \hline
\end{tabular}
\label{tab:smear}

\end{table}

\section{Tagging algorithm}
\label{sec:tag}

The tagging algorithm used for this study is based on the algorithm described in
Ref.~\cite{LHCb:2015tna}. %
The LHCb algorithm constructs secondary vertices (SVs) within jets and uses two
boosted decision tree (BDT) classifiers to identify vertices from light-, $c$-,
and $b$-hadron decays. %
One BDT is trained to distinguish heavy-flavor SVs from light-hadron SVs, and
the other is trained to distinguish between $b$ and $c$ SVs. %
For this study, the LHCb SV reconstruction algorithm was adapted to the
simulated EIC data. %
Heavy-vs.-light and $b$-vs.-$c$ BDTs were trained for a hypothetical EIC
detector using variables similar to those used to train the LHCb BDTs. %

First, displaced pseudo-reconstructed charged particles are combined to form
two-track SVs. %
Charged particle displacement is characterized by $\chi^2_{\rm DCA}$, which is
defined as in Eqn.~\ref{eqn:dcachi2} but with distances calculated with respect
to the smeared PV position instead of the true interaction point. %
Charged particles are considered displaced if \mbox{$\chi^2_{\rm DCA}>16$}. %
Pairs of displaced charged particles with a distance of closest approach to one
another less than $0.2~{\rm mm}$ are combined to form two-track SVs. %
Next, pairs of two-track SVs that share a track are combined to form three-track
SVs. %
Only SVs with \mbox{$0.4<m<5.3~{\rm GeV}$} are considered for merging, where $m$
is the SV mass calculated assuming the charged pion mass for each of the
constituent tracks. %
This merging process is repeated until no SVs share tracks. %
The resulting SVs can consist of any number of tracks. %

To suppress contributions from strange particle decays, two-track SVs are
required to have \mbox{$m>0.6~{\rm GeV}$}. %
This requirement removes both \mbox{$K^0_{\rm s}\to\pi^+\pi^-$} and
\mbox{$\Lambda\to p\pi^-$} decays. %
Events containing at least one SV passing these requirements are considered
tagged. %
If an event contains multiple SVs, the SV with the largest $p_{\rm T}$ is used
for further classification. %

\begin{figure*}
    {\centering
    \includegraphics[width=0.24\textwidth]{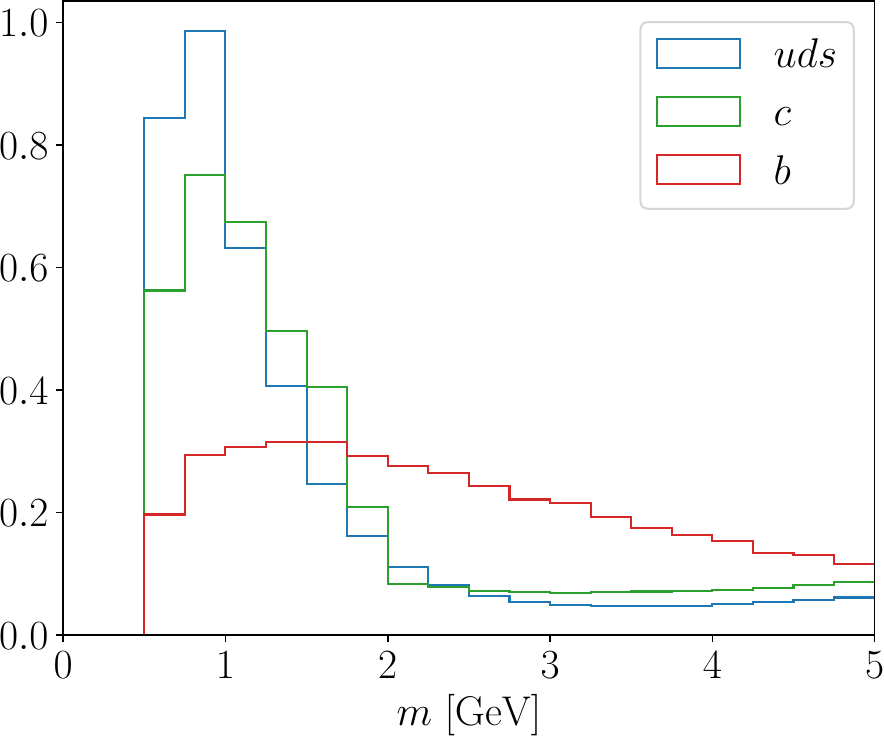}
    \includegraphics[width=0.24\textwidth]{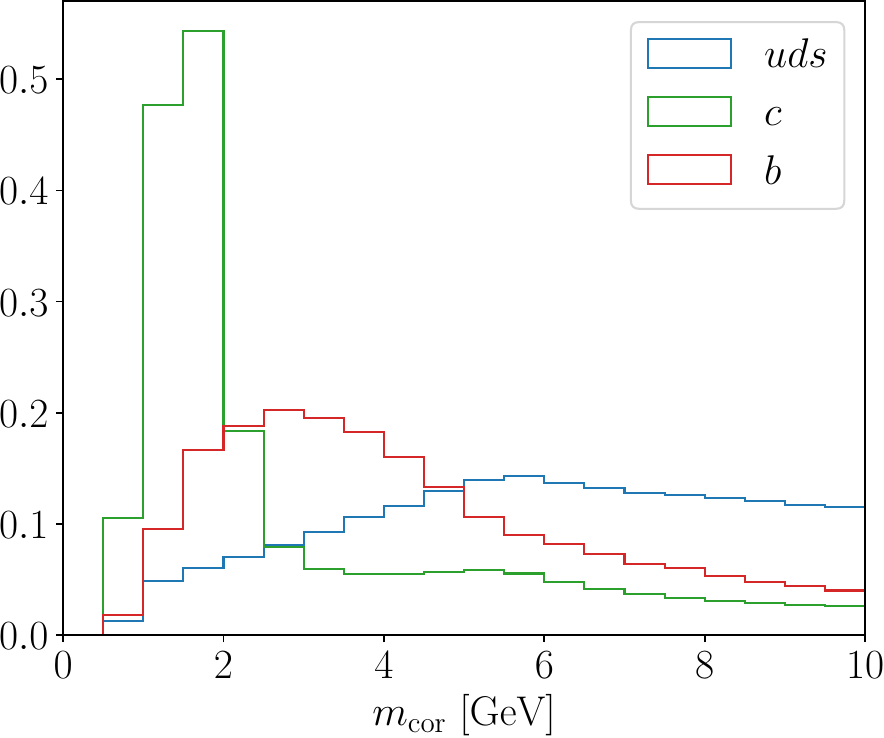}
    \includegraphics[width=0.24\textwidth]{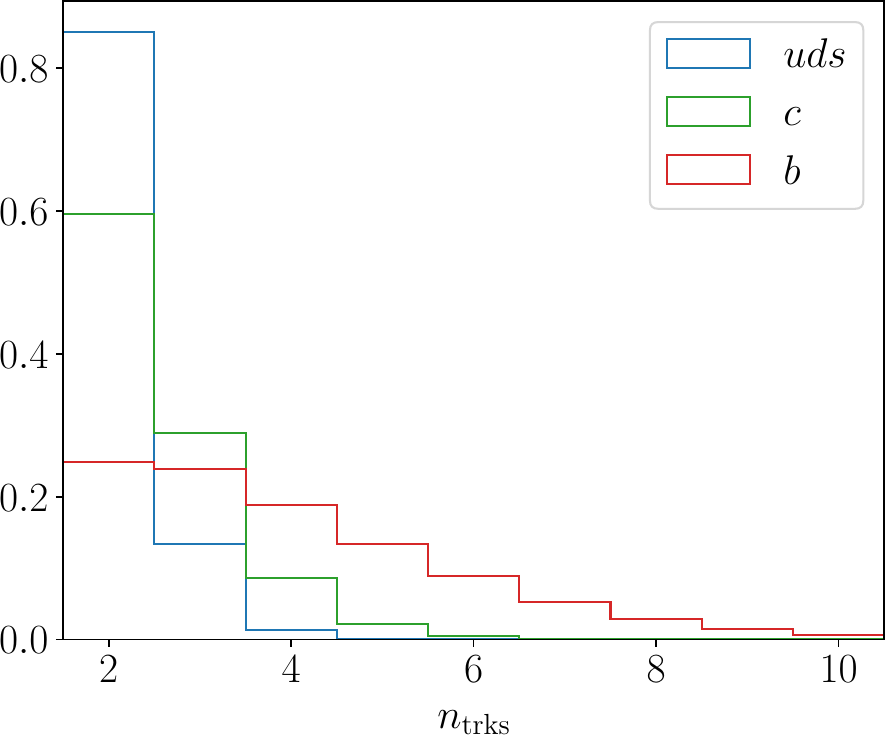}
    \includegraphics[width=0.24\textwidth]{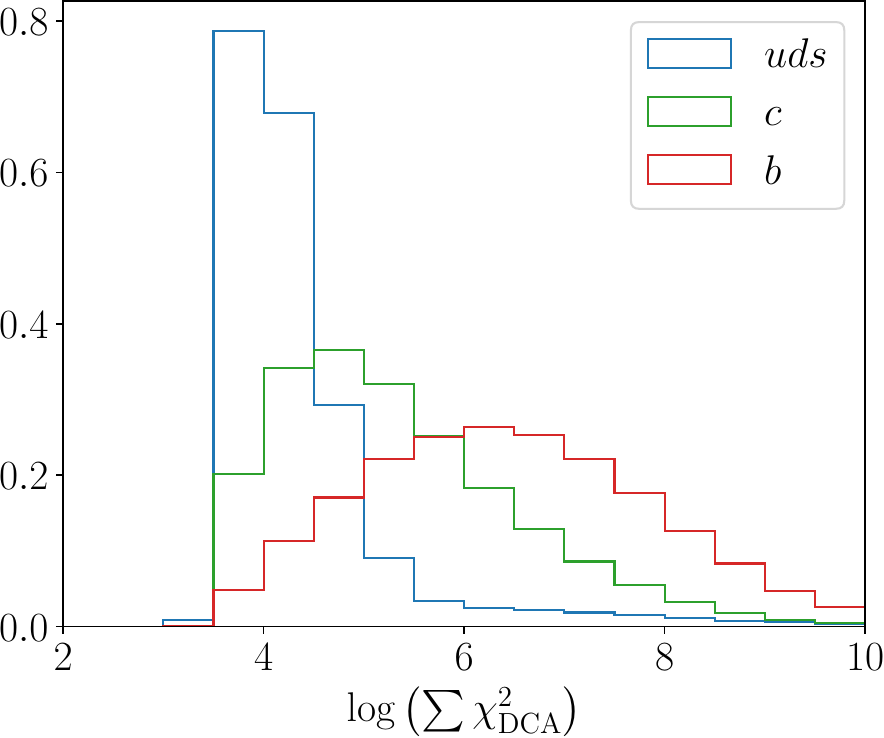}
    
    } 
\caption{Distributions of variables used for BDT training from the
\mbox{$\sqrt{s}=63~{\rm GeV}$} simulated data sample. Each distribution is normalized
to unit area.}
\label{fig:train}
    
\end{figure*}

Tagged events are classified using a pair of BDT classifiers. %
The first BDT is trained to distinguish heavy-flavor events from $uds$ events
(${\rm BDT}_{bc|uds}$), and the second is trained to distinguish between $b$
and $c$ events (${\rm BDT}_{b|c}$). %
The BDTs use four variables characterizing the SV tag. %
These include the mass of the SV ($m$), the number of tracks used to construct
the SV ($n_{\rm trks}$), and the sum of the $\chi^2_{\rm DCA}$ of the
constituent tracks. %
In addition, the BDTs use the corrected mass of the SV, which is given by
\begin{equation}
    m_{\rm cor}=\sqrt{m^2+p_{\perp}^2} + p_{\perp},
\end{equation}
where $p_{\perp}$ is the component of the SV momentum perpendicular to its
flight direction\,\cite{Williams:2011aza,SLD:1997ihw}. %
These variables are chosen because they depend only on the topological
properties of the SV and do not depend on the full SV covariance matrix, which
is difficult to estimate without a realistic detector simulation and
reconstruction algorithms. %

The distributions of the BDT input variables are shown in Fig.~\ref{fig:train}
for the \mbox{$\sqrt{s}=63~{\rm GeV}$} beam configuration. %
Bottom hadrons are more massive and produce more final-state particles than $c$
hadrons, which results in the observed hierarchies in $m$ and $n_{\rm trks}$. %
They also have a longer lifetime than $c$ and light hadrons and consequently
have larger $\sum\chi^2_{\rm DCA}$. %
The corrected mass is particularly powerful for identifying $c$ events because
$c$ hadrons typically decay at a single vertex. %
These decays produce a $m_{\rm cor}$ peak near the mass of the $D$ meson. %
Bottom hadrons produce more complex decay topologies and a consequently broader
$m_{\rm cor}$ distribution than that of charm hadrons. %
SVs in $uds$ events are made up of combinations of poorly-reconstructed prompt
tracks. %
The momenta of these combinations can point far from the PV and produce large
corrected masses. %

\begin{figure}

{\centering
\includegraphics[width=0.45\textwidth]{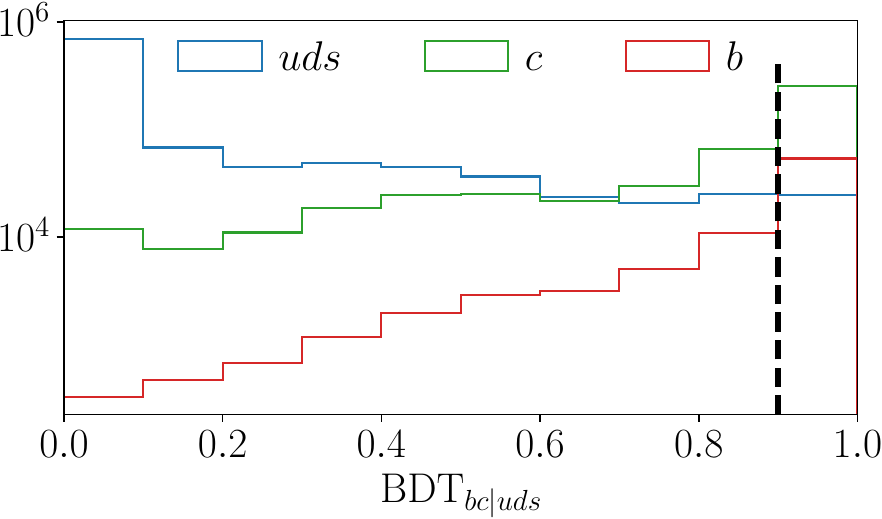}

\includegraphics[width=0.45\textwidth]{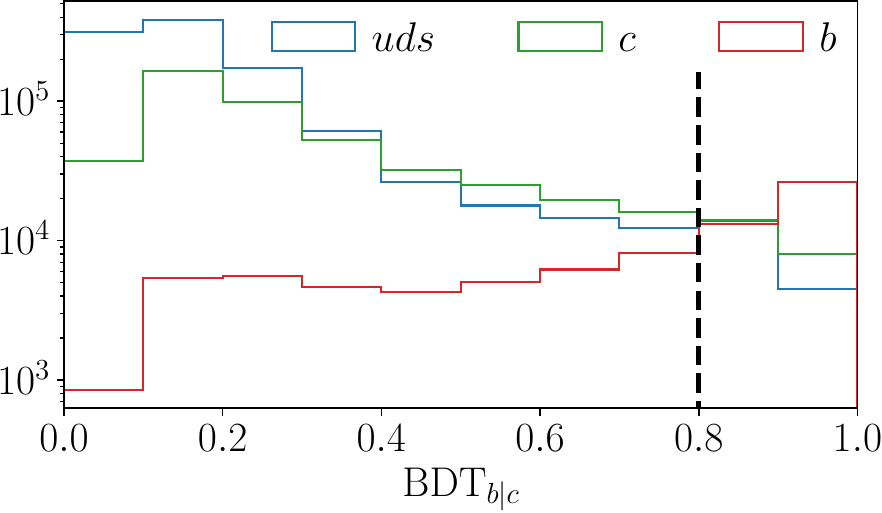}

} \caption{Response distributions for (top) ${\rm BDT}_{bc|uds}$ and (bottom)
${\rm BDT}_{b|c}$ from \mbox{$\sqrt{s}=63~{\rm GeV}$} simulation with expected
relative normalizations. The dashed lines show the low edges of the signal
regions.}
\label{fig:bdt}

\end{figure}

\begin{figure}
{\centering
\includegraphics[width=0.48\textwidth]{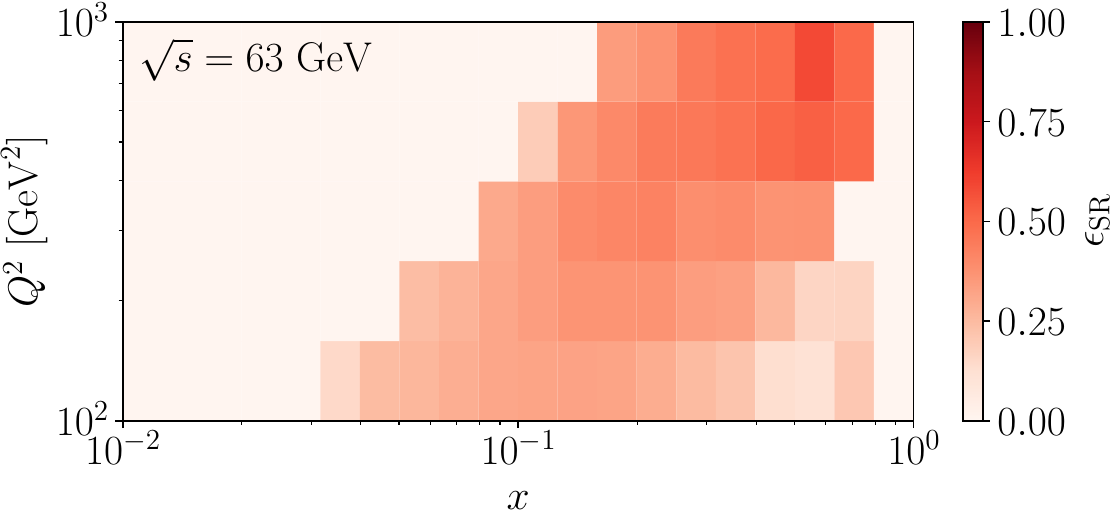}

} 
\caption{The signal region tagging efficiency determined from
$\sqrt{s}=63~{\rm GeV}$ simulation. Kinematically forbidden regions are given an
efficiency of zero.}
\label{fig:effsr}
    
\end{figure}

The BDT response distributions are shown in Fig.~\ref{fig:bdt}. %
In an analysis of real data, the composition of the tagged sample could be
determined using a two-dimensional template fit to these
distributions\,\cite{LHCb:2015bwt,LHCb:2015nta}. %
For this study, the region \mbox{${\rm BDT}_{bc|uds}>0.9$} and \mbox{${\rm
BDT}_{b|c}>0.8$} was defined as the signal region (SR) for the purpose of
estimating statistical uncertainties. %
The signal region tagging efficiency $\epsilon_{\rm SR}$, defined as the
probability that an event is tagged and the SV falls in the BDT SR, is shown in
Fig.~\ref{fig:effsr} for the \mbox{$\sqrt{s}=63~{\rm GeV}$} configuration. %
The tagging efficiency ranges from $30\mbox{--}40\%$ in most kinematically
allowed bins and approaches $60\%$ at high $Q^2$. %
This efficiency is consistent with the $b$-jet tagging efficiency observed by
LHCb, which approached $60\%$ at high jet $p_{\rm T}$. %
Charm events have a signal-region mistag probability of around $1\%$, while
$uds$ events have a mistag probability of around $10^{-4}$. %
While $uds$ events are the largest background overall, their contribution to the
signal region is small. %
The fast simulation used in this study does not include non-Gaussian
misreconstruction effects or secondary particle production from material
interactions. %
Both of these effects will create additional SVs in $uds$ events, but these SVs
should still be distinguishable from heavy flavor decays and are expected to
make a small contribution to the signal region\,\cite{LHCb:2015tna}. %

\section{Intrinsic bottom}
\label{sec:ib}

The $b\overline{b}$ production cross section in unpolarized neutral-current DIS
in the kinematic region studied here is given by
\begin{equation}
\frac{{\rm d}\sigma^{b{\overline{b}}}}{{\rm d}x{\rm d}Q^2}=\frac{2\pi\alpha^2Y_+}{xQ^4}\left(F^{b\overline{b}}_2(x,Q^2)-\frac{y^2}{Y_+}F^{b\overline{b}}_L(x,Q^2)\right),
\end{equation}
where $\alpha$ is the fine structure constant, \mbox{$Y_+=1+(1-y)^2$}, and
$F^{b\overline{b}}_2$ and $F^{b\overline{b}}_L$ are the $b$-quark contributions
to the proton structure functions\,\cite{H1:2018flt}. %
DIS experiments typically report a reduced cross section given by
\begin{equation}
    \sigma^{b\overline{b}}_r(x,Q^2)=F^{b\overline{b}}_2(x,Q^2)-\frac{y^2}{Y_+}F^{b\overline{b}}_L(x,Q^2).
\end{equation}
For the relatively small values of $y$ considered in this study,
$\sigma^{b\overline{b}}_r$ is determined primarily by $F^{b\overline{b}}_2$. At
leading order (LO) in the strong coupling constant $\alpha_{\rm s}$,
$F^{b\overline{b}}_2$ is proportional to the sum of $b$ and $\overline{b}$ PDFs.

The estimate of the IB contribution to $\sigma^{b\overline{b}}_r$ is based on
two observations. First, the intrinsic heavy quark PDFs evolve approximately
independently from the other PDFs\,\cite{Lyonnet:2015dca}. %
This means that the IC contribution to the charm PDF can be approximated as
\mbox{$c_{0+{\rm IC}}(x,Q^2)-c_0(x,Q^2)$}, where $c_0$ is the charm PDF from a
fit without IC and $c_{0+{\rm IC}}$ is from a fit that includes IC. %
The intrinsic $b$ PDF can then be estimated as
\begin{equation}
    b_{\rm IB}(x,Q^2)=\frac{m_c^2}{m_b^2}\left(c_{0+{\rm IC}}(x,Q^2)-c_0(x,Q^2)\right).
    \label{eqn:ib}
\end{equation}
Second, the dominant contribution from intrinsic heavy quarks to the reduced
cross section is from the LO contribution to $F_2^{q\overline{q}}$. %
As a result,
\begin{equation}
    \sigma^{b\overline{b}}_{r,{\rm IB}}(x,Q^2)\approx\sigma^{b\overline{b}}_{r{\rm no\mbox{-}IB}}(x,Q^2)+2e_{b}^2xb_{\rm IB}(x,Q^2),
\end{equation}
where $\sigma^{b\overline{b}}_{r{\rm no\mbox{-}IB}}$ is the reduced cross
section section assuming no IB, and $e_b$ is the electric charge of the $b$
quark. The factor of two in front of the IB term accounts for the $\overline{b}$
contribution, assuming the $b$ and $\overline{b}$ PDFs are symmetric. Applying
this strategy to calculate $\sigma^{c\overline{c}}_{r{\rm IC}}$ reproduces the
full next-to-next-to-leading order (NNLO) result to within about 10\% in the
kinematic region covered by this study, which is sufficiently accurate for the
sensitivity estimates performed here. Consequently, the IB contribution can be
estimated using only $c_0$, $c_{0+{\rm IC}}$, and $\sigma^{b\overline{b}}_{r{\rm
no\mbox{-}IB}}$.

The no-IB cross sections for $b$, $c$, and $uds$ events were calculated at NNLO
in $\alpha_{\rm s}$ using the Yadism package\,\cite{Candido:2024rkr}. %
The calculations were performed using the zero-mass variable flavor number
scheme (ZM-VFNS) and the CT18NNLO PDF set, which was accessed using
LHAPDF\,\cite{Hou:2019efy,Buckley:2014ana}. %
The no-IC charm PDF $c_0$ is taken from CT18NNLO, and $c_{0+{\rm IC}}$ was taken
from CT18FC\,\cite{Guzzi:2022rca}. %
CT18FC includes IC using the LFQCD-inspired model of Ref.~\cite{Brodsky:1980pb}
with $\left<x\right>_{\rm IC}\approx0.5\%$. %
It should be noted that the IC PDF from CT18FC is smaller than that from other
global analyses, and IC normalizations almost three times larger than that used for
this study are allowed within the CT18FC $68\%$ confidence interval. %
Furthermore, the $m_c^2/m_b^2$ IB scaling is unconfirmed. %
IB with an order-of-magnitude larger overall normalization has not been excluded
by data\cite{Lyonnet:2015dca}. %
In this sense, the IB model used in this study is conservative. %

The cross sections are used to calculate expected yields from one year of data
taking in each beam configuration according to the integrated luminosities given
in Refs.~\cite{Armesto:2023hnw} and \cite{Cerci:2023uhu}, which are reproduced
in Table~\ref{tab:lumi}. %
Signal-region tagging efficiencies for $b$, $c$, and $uds$ events were
calculated for each beam configuration as described in Sec.~\ref{sec:tag} and
were used to calculate expected tagged yields. %
The tagged yields were then used to determine signal significance and expected
statistical uncertainties. %
The IC contribution to the $c\overline{c}$ cross section is included as
background in the IB predictions. %

\begin{table}
    \caption{Expected annual integrated luminosities for various EIC beam configurations.}
    \begin{tabular}{c c}
        \hline
        $\sqrt{s}~[{\rm GeV}]$ & $\mathcal{L}_{\rm int}/{\rm year}~[{\rm fb}^{-1}]$\\
        \hline
        $45$ & 61.0\\
        $63$ & 79.0\\
        $105$ & 100.0\\
        $141$ & 15.4\\
        \hline
    \end{tabular}
    \label{tab:lumi}
\end{table}

\begin{figure*}
    {\centering
    \includegraphics[width=\textwidth]{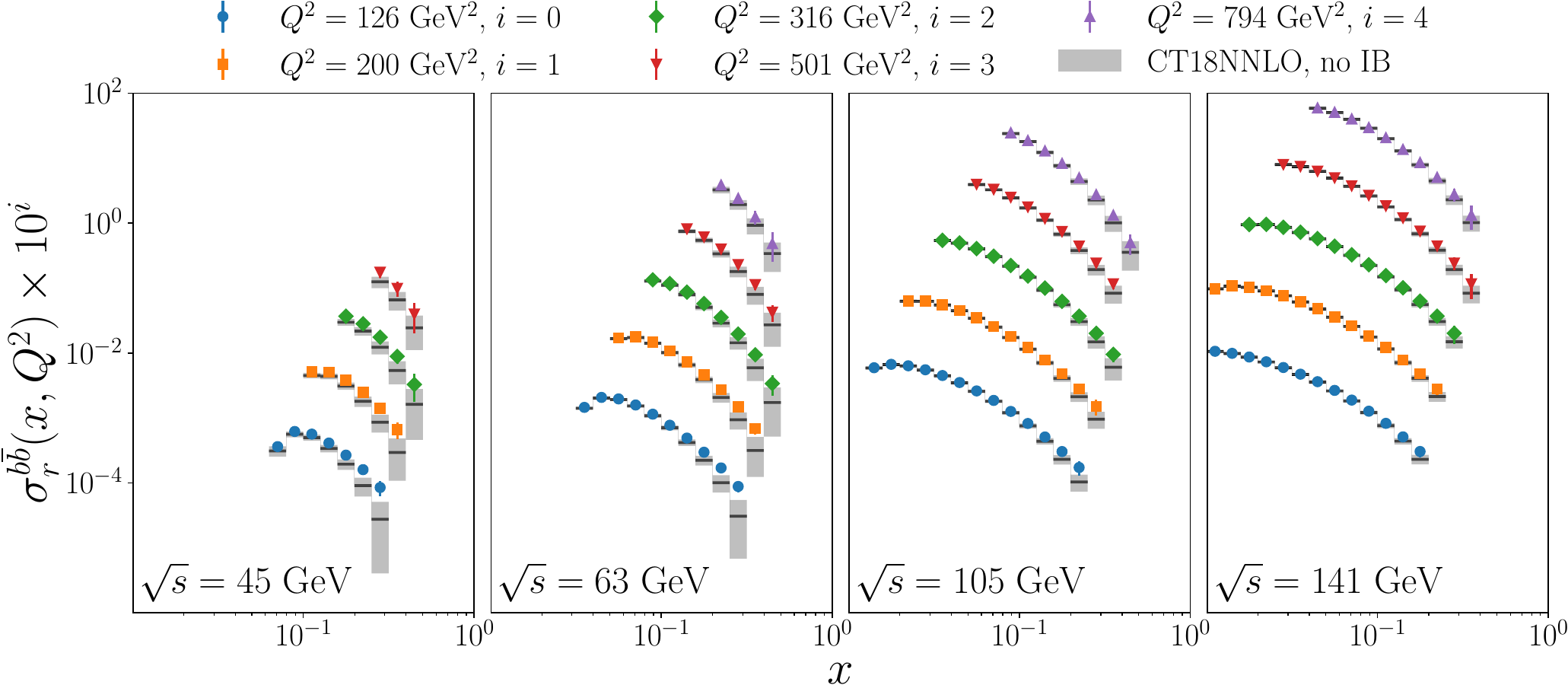}
    
    }

\caption{ Reduced cross section predictions. The points show predictions with
    IB, and the errorbars show expected statistical uncertainties. The black
    lines show predictions without IB, and the shaded boxes show the $68\%$
    confidence-level PDF uncertainties. For visual clarity, the result in each
    $Q^2$ bin is offset by a factor $10^i$, shown in the legend.}

\label{fig:rxsec}
    
\end{figure*}

\begin{figure*}
{\centering
\includegraphics[width=\textwidth]{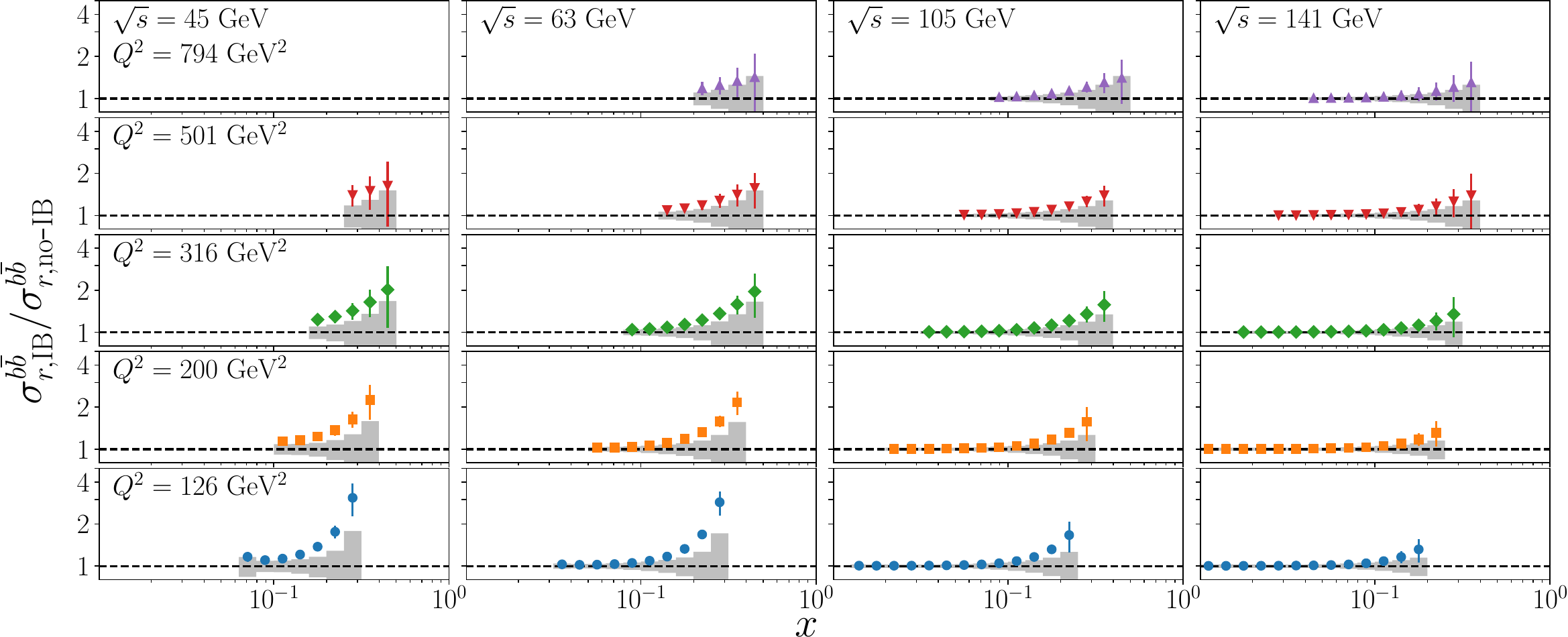}

} \caption{Ratio of the $b\overline{b}$ reduced cross section with IB to the no
IB case. The shaded regions show the $68\%$ confidence interval PDF
uncertainties for the no-IB case.}
\label{fig:rxsec_ratios}

\end{figure*}

The $\sigma_{r,{\rm IB}}^{b\overline{b}}$ results are shown in
Fig.~\ref{fig:rxsec}. %
To better illustrate the estimated sensitivity to IB, the ratio of the IB
results to the baseline are shown in Fig.~\ref{fig:rxsec_ratios}. %
IB produces an enhancement of up to a factor of $3$ in the valence region. %
The enhancement is most pronounced at low $Q^2$, where the contribution from
perturbative $b$ is smallest. %
In most of the kinematic bins where IB has a significant effect, the $b$ PDF
uncertainties are much larger than the expected statistical uncertainties. %
Because the no-IB PDF is determined entirely from gluon splitting via DGLAP
evolution, these PDF uncertainties reflect uncertainties in the gluon
PDF at high $x$. %
Consequently, the EIC's sensitivity to IB will depend in part on future
constraints on the high-$x$ gluon PDF from both the EIC and the LHC. %

Using LHCb's experience as a guide, the largest systematic uncertainties for
measurements using this tagging algorithm are likely to arise from the tagging
efficiency determination and the BDT template calibration. %
The LHCb experiment was able to measure its jet tagging efficiency in data to
within about 10\% and calibrate its templates using dijet calibration
samples\,\cite{LHCb:2021dlw,LHCb:2015tna}. %
A similar calibration is possible for $c\overline{c}$ events at the EIC using
SV-tagged events containing a fully reconstructed \mbox{$D^0\to K^-\pi^+$} decay
with a large separation in azimuthal angle $\phi$ from the tagging SV. %
The $b$ tagging performance can be studied using events containing two $b$-like
tags with large separations in $\phi$. %
Ultimately, because the same templates are used for efficiency determinations
and signal yield extraction, these uncertainties partially cancel in actual
measurements. %
Furthermore, the remaining uncertainty will likely be highly correlated across
kinematic bins and should only mildly affect sensitivity to IB. %
Measurements of heavy flavor production by the H1 and ZEUS collaborations using
topological tagging also found that the dominant systematic uncertainties were
highly correlated between data points\,\cite{H1:2009uwa,ZEUS:2014wft}. %

The search for IB will also be complicated by the handling of the $b$-quark
mass in the $b$ PDF evolution. %
The $b$-quark pole mass is typically used as a starting scale for generating $b$ quarks
perturbatively and is anticorrelated with the $b$ PDF. %
Variations of $m_b$ within its uncertainties can produce changes in the $b$ PDF
comparable to the PDF fit uncertainties in the valence
region\,\cite{Cridge:2021qfd,Campbell:2021qgd}. %
Varying $m_b$ has a much larger effect at low $x$, however, and data over the
broad $x$ range studied here would provide strong constraints on both IB and
$m_b$ simultaneously. %
This data would also provide powerful tests of the heavy-quark scheme used in
structure function calculations\,\cite{Accardi:2016ndt}. %

\section{Conclusions}
\label{sec:conc}

Topological $b$ tagging has proven to be a powerful tool for studying QCD in
high-energy hadron collisions, and this work demonstrates that these methods are
directly applicable to the EIC. %
The tagging strategy described here has a wide range of potential applications
in electron-proton and electron-nucleus scattering. %
This algorithm could be used to tag heavy-flavor jets, which can be used to
study both the structure of nuclei and the hadronization
process\,\cite{Arratia:2020azl,Li:2021gjw}. %
It could also be used to study heavy dihadron angular correlations, which
provide sensitivity to gluon transverse momentum
distributions\cite{delCastillo:2021znl,Dong:2022xbd}. %
This tagging strategy can also be used to efficiently tag charm events,
potentially expanding the kinematic reach of charm production measurements at
the EIC. %

This paper also presents the first study of the EIC's ability to probe the
$b$-quark PDF and its sensitivity to intrinsic bottom quarks. %
The EIC has the potential to observe intrinsic bottom at levels expected from
recent global analyses of intrinsic charm in the proton. %
The observation of intrinsic bottom quarks is crucial for understanding the
origin of intrinsic heavy quarks, including possible nonperturbative processes
that produce heavy quarks in protons and nuclei. %
This paper presents the first strategy for observing intrinsic bottom in the
near future. %

\section*{Acknowledgements}

The author thanks Matthew Durham, Philip Ilten, Paul Newman, Brian Page, Michael
Sokoloff, and Michael Williams for helpful discussions and feedback. %
This material is based upon work supported by the National Science Foundation
under Grant No. PHY-2208983. %
Any opinions, findings, and conclusions or recommendations expressed in this
material are those of the author and do not necessarily reflect the views of the
National Science Foundation. %

\clearpage

\bibliography{main}

\begin{thebibliography}{42}%
\makeatletter
\providecommand \@ifxundefined [1]{%
 \@ifx{#1\undefined}
}%
\providecommand \@ifnum [1]{%
 \ifnum #1\expandafter \@firstoftwo
 \else \expandafter \@secondoftwo
 \fi
}%
\providecommand \@ifx [1]{%
 \ifx #1\expandafter \@firstoftwo
 \else \expandafter \@secondoftwo
 \fi
}%
\providecommand \natexlab [1]{#1}%
\providecommand \enquote  [1]{``#1''}%
\providecommand \bibnamefont  [1]{#1}%
\providecommand \bibfnamefont [1]{#1}%
\providecommand \citenamefont [1]{#1}%
\providecommand \href@noop [0]{\@secondoftwo}%
\providecommand \href [0]{\begingroup \@sanitize@url \@href}%
\providecommand \@href[1]{\@@startlink{#1}\@@href}%
\providecommand \@@href[1]{\endgroup#1\@@endlink}%
\providecommand \@sanitize@url [0]{\catcode `\\12\catcode `\$12\catcode
  `\&12\catcode `\#12\catcode `\^12\catcode `\_12\catcode `\%12\relax}%
\providecommand \@@startlink[1]{}%
\providecommand \@@endlink[0]{}%
\providecommand \url  [0]{\begingroup\@sanitize@url \@url }%
\providecommand \@url [1]{\endgroup\@href {#1}{\urlprefix }}%
\providecommand \urlprefix  [0]{URL }%
\providecommand \Eprint [0]{\href }%
\providecommand \doibase [0]{https://doi.org/}%
\providecommand \selectlanguage [0]{\@gobble}%
\providecommand \bibinfo  [0]{\@secondoftwo}%
\providecommand \bibfield  [0]{\@secondoftwo}%
\providecommand \translation [1]{[#1]}%
\providecommand \BibitemOpen [0]{}%
\providecommand \bibitemStop [0]{}%
\providecommand \bibitemNoStop [0]{.\EOS\space}%
\providecommand \EOS [0]{\spacefactor3000\relax}%
\providecommand \BibitemShut  [1]{\csname bibitem#1\endcsname}%
\let\auto@bib@innerbib\@empty
\bibitem [{\citenamefont {Brodsky}\ \emph {et~al.}(1980)\citenamefont
  {Brodsky}, \citenamefont {Hoyer}, \citenamefont {Peterson},\ and\
  \citenamefont {Sakai}}]{Brodsky:1980pb}%
  \BibitemOpen
  \bibfield  {author} {\bibinfo {author} {\bibfnamefont {S.~J.}\ \bibnamefont
  {Brodsky}}, \bibinfo {author} {\bibfnamefont {P.}~\bibnamefont {Hoyer}},
  \bibinfo {author} {\bibfnamefont {C.}~\bibnamefont {Peterson}},\ and\
  \bibinfo {author} {\bibfnamefont {N.}~\bibnamefont {Sakai}},\ }
  \href
  {https://doi.org/10.1016/0370-2693(80)90364-0} {\bibfield  {journal}
  {\bibinfo  {journal} {Phys. Lett.}\ }\textbf {\bibinfo {volume} {B93}},\
  \bibinfo {pages} {451} (\bibinfo {year} {1980})}\BibitemShut {NoStop}%
\bibitem [{\citenamefont {Hobbs}\ \emph {et~al.}(2014)\citenamefont {Hobbs},
  \citenamefont {Londergan},\ and\ \citenamefont
  {Melnitchouk}}]{Hobbs:2013bia}%
  \BibitemOpen
  \bibfield  {author} {\bibinfo {author} {\bibfnamefont {T.~J.}\ \bibnamefont
  {Hobbs}}, \bibinfo {author} {\bibfnamefont {J.~T.}\ \bibnamefont
  {Londergan}},\ and\ \bibinfo {author} {\bibfnamefont {W.}~\bibnamefont
  {Melnitchouk}},\ }
  \href
  {https://doi.org/10.1103/PhysRevD.89.074008} {\bibfield  {journal} {\bibinfo
  {journal} {Phys. Rev.}\ }\textbf {\bibinfo {volume} {D89}},\ \bibinfo
  {pages} {074008} (\bibinfo {year} {2014})},\ \Eprint
  {https://arxiv.org/abs/1311.1578} {arXiv:1311.1578 [hep-ph]} \BibitemShut
  {NoStop}%
\bibitem [{\citenamefont {Guzzi}\ \emph {et~al.}(2023)\citenamefont {Guzzi},
  \citenamefont {Hobbs}, \citenamefont {Xie}, \citenamefont {Huston},
  \citenamefont {Nadolsky},\ and\ \citenamefont {Yuan}}]{Guzzi:2022rca}%
  \BibitemOpen
  \bibfield  {author} {\bibinfo {author} {\bibfnamefont {M.}~\bibnamefont
  {Guzzi}}, \bibinfo {author} {\bibfnamefont {T.~J.}\ \bibnamefont {Hobbs}},
  \bibinfo {author} {\bibfnamefont {K.}~\bibnamefont {Xie}}, \bibinfo {author}
  {\bibfnamefont {J.}~\bibnamefont {Huston}}, \bibinfo {author} {\bibfnamefont
  {P.}~\bibnamefont {Nadolsky}},\ and\ \bibinfo {author} {\bibfnamefont
  {C.~P.}\ \bibnamefont {Yuan}},\ }
  \href {https://doi.org/10.1016/j.physletb.2023.137975} {\bibfield  {journal}
  {\bibinfo  {journal} {Phys. Lett.}\ }\textbf {\bibinfo {volume} {B843}},\
  \bibinfo {pages} {137975} (\bibinfo {year} {2023})},\ \Eprint
  {https://arxiv.org/abs/2211.01387} {arXiv:2211.01387 [hep-ph]} \BibitemShut
  {NoStop}%
\bibitem [{\citenamefont {Hou}\ \emph {et~al.}(2021)\citenamefont {Hou} \emph
  {et~al.}}]{Hou:2019efy}%
  \BibitemOpen
  \bibfield  {author} {\bibinfo {author} {\bibfnamefont {T.-J.}\ \bibnamefont
  {Hou}} \emph {et~al.},\ }
  \href {https://doi.org/10.1103/PhysRevD.103.014013} {\bibfield
  {journal} {\bibinfo  {journal} {Phys. Rev.}\ }\textbf {\bibinfo {volume}
  {D103}},\ \bibinfo {pages} {014013} (\bibinfo {year} {2021})},\ \Eprint
  {https://arxiv.org/abs/1912.10053} {arXiv:1912.10053 [hep-ph]} \BibitemShut
  {NoStop}%
\bibitem [{\citenamefont {Aubert}\ \emph {et~al.}(1983)\citenamefont {Aubert}
  \emph {et~al.}}]{EuropeanMuon:1982xfn}%
  \BibitemOpen
  \bibfield  {author} {\bibinfo {author} {\bibfnamefont {J.~J.}\ \bibnamefont
  {Aubert}} \emph {et~al.} (\bibinfo {collaboration} {European Muon}),\
  }\href
  {https://doi.org/10.1016/0550-3213(83)90174-8} {\bibfield  {journal}
  {\bibinfo  {journal} {Nucl. Phys.}\ }\textbf {\bibinfo {volume} {B213}},\
  \bibinfo {pages} {31} (\bibinfo {year} {1983})}\BibitemShut {NoStop}%
\bibitem [{\citenamefont {Aaij}\ \emph
  {et~al.}(2022{\natexlab{a}})\citenamefont {Aaij} \emph
  {et~al.}}]{LHCb:2021stx}%
  \BibitemOpen
  \bibfield  {author} {\bibinfo {author} {\bibfnamefont {R.}~\bibnamefont
  {Aaij}} \emph {et~al.} (\bibinfo {collaboration} {LHCb}),\ }\href
  {https://doi.org/10.1103/PhysRevLett.128.082001} {\bibfield  {journal}
  {\bibinfo  {journal} {Phys. Rev. Lett.}\ }\textbf {\bibinfo {volume} {128}},\
  \bibinfo {pages} {082001} (\bibinfo {year} {2022}{\natexlab{a}})},\ \Eprint
  {https://arxiv.org/abs/2109.08084} {arXiv:2109.08084 [hep-ex]} \BibitemShut
  {NoStop}%
\bibitem [{\citenamefont {Boettcher}\ \emph {et~al.}(2016)\citenamefont
  {Boettcher}, \citenamefont {Ilten},\ and\ \citenamefont
  {Williams}}]{Boettcher:2015sqn}%
  \BibitemOpen
  \bibfield  {author} {\bibinfo {author} {\bibfnamefont {T.}~\bibnamefont
  {Boettcher}}, \bibinfo {author} {\bibfnamefont {P.}~\bibnamefont {Ilten}},\
  and\ \bibinfo {author} {\bibfnamefont {M.}~\bibnamefont {Williams}},\
  }\href {https://doi.org/10.1103/PhysRevD.93.074008}
  {\bibfield  {journal} {\bibinfo  {journal} {Phys. Rev.}\ }\textbf {\bibinfo
  {volume} {D93}},\ \bibinfo {pages} {074008} (\bibinfo {year} {2016})},\
  \Eprint {https://arxiv.org/abs/1512.06666} {arXiv:1512.06666 [hep-ph]}
  \BibitemShut {NoStop}%
\bibitem [{\citenamefont {Aaij}\ \emph {et~al.}(2019)\citenamefont {Aaij} \emph
  {et~al.}}]{LHCb:2018jry}%
  \BibitemOpen
  \bibfield  {author} {\bibinfo {author} {\bibfnamefont {R.}~\bibnamefont
  {Aaij}} \emph {et~al.} (\bibinfo {collaboration} {LHCb}),\ }\href
  {https://doi.org/10.1103/PhysRevLett.122.132002} {\bibfield  {journal}
  {\bibinfo  {journal} {Phys. Rev. Lett.}\ }\textbf {\bibinfo {volume} {122}},\
  \bibinfo {pages} {132002} (\bibinfo {year} {2019})},\ \Eprint
  {https://arxiv.org/abs/1810.07907} {arXiv:1810.07907 [hep-ex]} \BibitemShut
  {NoStop}%
\bibitem [{\citenamefont {Aaij}\ \emph {et~al.}(2023)\citenamefont {Aaij} \emph
  {et~al.}}]{LHCb:2022cul}%
  \BibitemOpen
  \bibfield  {author} {\bibinfo {author} {\bibfnamefont {R.}~\bibnamefont
  {Aaij}} \emph {et~al.} (\bibinfo {collaboration} {LHCb}),\ }\href
  {https://doi.org/10.1140/epjc/s10052-023-11641-5} {\bibfield  {journal}
  {\bibinfo  {journal} {Eur. Phys. J.}\ }\textbf {\bibinfo {volume} {C83}},\
  \bibinfo {pages} {541} (\bibinfo {year} {2023})},\ \Eprint
  {https://arxiv.org/abs/2211.11633} {arXiv:2211.11633 [hep-ex]} \BibitemShut
  {NoStop}%
\bibitem [{\citenamefont {Ball}\ \emph {et~al.}(2022)\citenamefont {Ball},
  \citenamefont {Candido}, \citenamefont {Cruz-Martinez}, \citenamefont
  {Forte}, \citenamefont {Giani}, \citenamefont {Hekhorn}, \citenamefont
  {Kudashkin}, \citenamefont {Magni},\ and\ \citenamefont
  {Rojo}}]{Ball:2022qks}%
  \BibitemOpen
  \bibfield  {author} {\bibinfo {author} {\bibfnamefont {R.~D.}\ \bibnamefont
  {Ball}}, \bibinfo {author} {\bibfnamefont {A.}~\bibnamefont {Candido}},
  \bibinfo {author} {\bibfnamefont {J.}~\bibnamefont {Cruz-Martinez}}, \bibinfo
  {author} {\bibfnamefont {S.}~\bibnamefont {Forte}}, \bibinfo {author}
  {\bibfnamefont {T.}~\bibnamefont {Giani}}, \bibinfo {author} {\bibfnamefont
  {F.}~\bibnamefont {Hekhorn}}, \bibinfo {author} {\bibfnamefont
  {K.}~\bibnamefont {Kudashkin}}, \bibinfo {author} {\bibfnamefont
  {G.}~\bibnamefont {Magni}},\ and\ \bibinfo {author} {\bibfnamefont
  {J.}~\bibnamefont {Rojo}} (\bibinfo {collaboration} {NNPDF}),\ }
  \href {https://doi.org/10.1038/s41586-022-04998-2} {\bibfield
  {journal} {\bibinfo  {journal} {Nature}\ }\textbf {\bibinfo {volume} {608}},\
  \bibinfo {pages} {483} (\bibinfo {year} {2022})},\ \Eprint
  {https://arxiv.org/abs/2208.08372} {arXiv:2208.08372 [hep-ph]} \BibitemShut
  {NoStop}%
\bibitem [{\citenamefont {Abdul~Khalek}\ \emph {et~al.}(2022)\citenamefont
  {Abdul~Khalek} \emph {et~al.}}]{AbdulKhalek:2021gbh}%
  \BibitemOpen
  \bibfield  {author} {\bibinfo {author} {\bibfnamefont {R.}~\bibnamefont
  {Abdul~Khalek}} \emph {et~al.},\ }\href
  {https://doi.org/10.1016/j.nuclphysa.2022.122447} {\bibfield  {journal}
  {\bibinfo  {journal} {Nucl. Phys.}\ }\textbf {\bibinfo {volume} {A1026}},\
  \bibinfo {pages} {122447} (\bibinfo {year} {2022})},\ \Eprint
  {https://arxiv.org/abs/2103.05419} {arXiv:2103.05419 [physics.ins-det]}
  \BibitemShut {NoStop}%
\bibitem [{\citenamefont {Ball}\ \emph {et~al.}(2023)\citenamefont {Ball},
  \citenamefont {Candido}, \citenamefont {Cruz-Martinez}, \citenamefont
  {Forte}, \citenamefont {Giani}, \citenamefont {Hekhorn}, \citenamefont
  {Magni}, \citenamefont {Nocera}, \citenamefont {Rojo},\ and\ \citenamefont
  {Stegeman}}]{NNPDF:2023tyk}%
  \BibitemOpen
  \bibfield  {author} {\bibinfo {author} {\bibfnamefont {R.~D.}\ \bibnamefont
  {Ball}}, \bibinfo {author} {\bibfnamefont {A.}~\bibnamefont {Candido}},
  \bibinfo {author} {\bibfnamefont {J.}~\bibnamefont {Cruz-Martinez}}, \bibinfo
  {author} {\bibfnamefont {S.}~\bibnamefont {Forte}}, \bibinfo {author}
  {\bibfnamefont {T.}~\bibnamefont {Giani}}, \bibinfo {author} {\bibfnamefont
  {F.}~\bibnamefont {Hekhorn}}, \bibinfo {author} {\bibfnamefont
  {G.}~\bibnamefont {Magni}}, \bibinfo {author} {\bibfnamefont {E.~R.}\
  \bibnamefont {Nocera}}, \bibinfo {author} {\bibfnamefont {J.}~\bibnamefont
  {Rojo}},\ and\ \bibinfo {author} {\bibfnamefont {R.}~\bibnamefont {Stegeman}}
  (\bibinfo {collaboration} {NNPDF}),\ }\href@noop {} {\  (\bibinfo {year} {2023})},\ \Eprint
  {https://arxiv.org/abs/2311.00743} {arXiv:2311.00743 [hep-ph]} \BibitemShut
  {NoStop}%
\bibitem [{\citenamefont {Kelsey}\ \emph {et~al.}(2021)\citenamefont {Kelsey},
  \citenamefont {Cruz-Torres}, \citenamefont {Dong}, \citenamefont {Ji},
  \citenamefont {Radhakrishnan},\ and\ \citenamefont
  {Sichtermann}}]{Kelsey:2021gpk}%
  \BibitemOpen
  \bibfield  {author} {\bibinfo {author} {\bibfnamefont {M.}~\bibnamefont
  {Kelsey}}, \bibinfo {author} {\bibfnamefont {R.}~\bibnamefont {Cruz-Torres}},
  \bibinfo {author} {\bibfnamefont {X.}~\bibnamefont {Dong}}, \bibinfo {author}
  {\bibfnamefont {Y.}~\bibnamefont {Ji}}, \bibinfo {author} {\bibfnamefont
  {S.}~\bibnamefont {Radhakrishnan}},\ and\ \bibinfo {author} {\bibfnamefont
  {E.}~\bibnamefont {Sichtermann}},\ }\href
  {https://doi.org/10.1103/PhysRevD.104.054002} {\bibfield  {journal} {\bibinfo
   {journal} {Phys. Rev.}\ }\textbf {\bibinfo {volume} {D104}},\ \bibinfo
  {pages} {054002} (\bibinfo {year} {2021})},\ \Eprint
  {https://arxiv.org/abs/2107.05632} {arXiv:2107.05632 [hep-ph]} \BibitemShut
  {NoStop}%
\bibitem [{\citenamefont {Brodsky}\ \emph {et~al.}(2015)\citenamefont
  {Brodsky}, \citenamefont {Kusina}, \citenamefont {Lyonnet}, \citenamefont
  {Schienbein}, \citenamefont {Spiesberger},\ and\ \citenamefont
  {Vogt}}]{Brodsky:2015fna}%
  \BibitemOpen
  \bibfield  {author} {\bibinfo {author} {\bibfnamefont {S.~J.}\ \bibnamefont
  {Brodsky}}, \bibinfo {author} {\bibfnamefont {A.}~\bibnamefont {Kusina}},
  \bibinfo {author} {\bibfnamefont {F.}~\bibnamefont {Lyonnet}}, \bibinfo
  {author} {\bibfnamefont {I.}~\bibnamefont {Schienbein}}, \bibinfo {author}
  {\bibfnamefont {H.}~\bibnamefont {Spiesberger}},\ and\ \bibinfo {author}
  {\bibfnamefont {R.}~\bibnamefont {Vogt}},\ }\href {https://doi.org/10.1155/2015/231547} {\bibfield  {journal} {\bibinfo
  {journal} {Adv. High Energy Phys.}\ }\textbf {\bibinfo {volume} {2015}},\
  \bibinfo {pages} {231547} (\bibinfo {year} {2015})},\ \Eprint
  {https://arxiv.org/abs/1504.06287} {arXiv:1504.06287 [hep-ph]} \BibitemShut
  {NoStop}%
\bibitem [{\citenamefont {Lyonnet}\ \emph {et~al.}(2015)\citenamefont
  {Lyonnet}, \citenamefont {Kusina}, \citenamefont {Je\v{z}o}, \citenamefont
  {Kovar\'\i{}k}, \citenamefont {Olness}, \citenamefont {Schienbein},\ and\
  \citenamefont {Yu}}]{Lyonnet:2015dca}%
  \BibitemOpen
  \bibfield  {author} {\bibinfo {author} {\bibfnamefont {F.}~\bibnamefont
  {Lyonnet}}, \bibinfo {author} {\bibfnamefont {A.}~\bibnamefont {Kusina}},
  \bibinfo {author} {\bibfnamefont {T.}~\bibnamefont {Je\v{z}o}}, \bibinfo
  {author} {\bibfnamefont {K.}~\bibnamefont {Kovar\'\i{}k}}, \bibinfo {author}
  {\bibfnamefont {F.}~\bibnamefont {Olness}}, \bibinfo {author} {\bibfnamefont
  {I.}~\bibnamefont {Schienbein}},\ and\ \bibinfo {author} {\bibfnamefont
  {J.-Y.}\ \bibnamefont {Yu}},\ }\href {https://doi.org/10.1007/JHEP07(2015)141} {\bibfield
  {journal} {\bibinfo  {journal} {JHEP}\ }\textbf {\bibinfo {volume} {07}},\
  \bibinfo {pages} {141}},\ \Eprint {https://arxiv.org/abs/1504.05156}
  {arXiv:1504.05156 [hep-ph]} \BibitemShut {NoStop}%
\bibitem [{\citenamefont {Workman}\ \emph {et~al.}(2022)\citenamefont {Workman}
  \emph {et~al.}}]{ParticleDataGroup:2022pth}%
  \BibitemOpen
  \bibfield  {author} {\bibinfo {author} {\bibfnamefont {R.~L.}\ \bibnamefont
  {Workman}} \emph {et~al.} (\bibinfo {collaboration} {Particle Data Group}),\
  }\href
  {https://doi.org/10.1093/ptep/ptac097} {\bibfield  {journal} {\bibinfo
  {journal} {PTEP}\ }\textbf {\bibinfo {volume} {2022}},\ \bibinfo {pages}
  {083C01} (\bibinfo {year} {2022})}\BibitemShut {NoStop}%
\bibitem [{\citenamefont {Abramowicz}\ \emph {et~al.}(2018)\citenamefont
  {Abramowicz} \emph {et~al.}}]{H1:2018flt}%
  \BibitemOpen
  \bibfield  {author} {\bibinfo {author} {\bibfnamefont {H.}~\bibnamefont
  {Abramowicz}} \emph {et~al.} (\bibinfo {collaboration} {H1, ZEUS}),\
  }\href
  {https://doi.org/10.1140/epjc/s10052-018-5848-3} {\bibfield  {journal}
  {\bibinfo  {journal} {Eur. Phys. J.}\ }\textbf {\bibinfo {volume} {C78}},\
  \bibinfo {pages} {473} (\bibinfo {year} {2018})},\ \Eprint
  {https://arxiv.org/abs/1804.01019} {arXiv:1804.01019 [hep-ex]} \BibitemShut
  {NoStop}%
\bibitem [{\citenamefont {Aaron}\ \emph {et~al.}(2010)\citenamefont {Aaron}
  \emph {et~al.}}]{H1:2009uwa}%
  \BibitemOpen
  \bibfield  {author} {\bibinfo {author} {\bibfnamefont {F.~D.}\ \bibnamefont
  {Aaron}} \emph {et~al.} (\bibinfo {collaboration} {H1}),\ }\href
  {https://doi.org/10.1140/epjc/s10052-009-1190-0} {\bibfield  {journal}
  {\bibinfo  {journal} {Eur. Phys. J.}\ }\textbf {\bibinfo {volume} {C65}},\
  \bibinfo {pages} {89} (\bibinfo {year} {2010})},\ \Eprint
  {https://arxiv.org/abs/0907.2643} {arXiv:0907.2643 [hep-ex]} \BibitemShut
  {NoStop}%
\bibitem [{\citenamefont {Abramowicz}\ \emph {et~al.}(2014)\citenamefont
  {Abramowicz} \emph {et~al.}}]{ZEUS:2014wft}%
  \BibitemOpen
  \bibfield  {author} {\bibinfo {author} {\bibfnamefont {H.}~\bibnamefont
  {Abramowicz}} \emph {et~al.} (\bibinfo {collaboration} {ZEUS}),\ }
  \href {https://doi.org/10.1007/JHEP09(2014)127} {\bibfield
  {journal} {\bibinfo  {journal} {JHEP}\ }\textbf {\bibinfo {volume} {09}},\
  \bibinfo {pages} {127}},\ \Eprint {https://arxiv.org/abs/1405.6915}
  {arXiv:1405.6915 [hep-ex]} \BibitemShut {NoStop}%
\bibitem [{\citenamefont {Aaij}\ \emph
  {et~al.}(2015{\natexlab{a}})\citenamefont {Aaij} \emph
  {et~al.}}]{LHCb:2015tna}%
  \BibitemOpen
  \bibfield  {author} {\bibinfo {author} {\bibfnamefont {R.}~\bibnamefont
  {Aaij}} \emph {et~al.} (\bibinfo {collaboration} {LHCb}),\ }
  \href {https://doi.org/10.1088/1748-0221/10/06/P06013} {\bibfield
  {journal} {\bibinfo  {journal} {JINST}\ }\textbf {\bibinfo {volume}
  {10}}\bibfield  {number} {\bibinfo  {number} { (06)},\ \bibinfo {pages}
  {P06013}},\ }\Eprint {https://arxiv.org/abs/1504.07670} {arXiv:1504.07670
  [hep-ex]} \BibitemShut {NoStop}%
\bibitem [{\citenamefont {Aaij}\ \emph
  {et~al.}(2022{\natexlab{b}})\citenamefont {Aaij} \emph
  {et~al.}}]{LHCb:2021dlw}%
  \BibitemOpen
  \bibfield  {author} {\bibinfo {author} {\bibfnamefont {R.}~\bibnamefont
  {Aaij}} \emph {et~al.} (\bibinfo {collaboration} {LHCb}),\ }\href
  {https://doi.org/10.1088/1748-0221/17/02/P02028} {\bibfield  {journal}
  {\bibinfo  {journal} {JINST}\ }\textbf {\bibinfo {volume} {17}}\bibfield
  {number} {\bibinfo  {number} { (02)},\ \bibinfo {pages} {P02028}},\ }\Eprint
  {https://arxiv.org/abs/2112.08435} {arXiv:2112.08435 [hep-ex]} \BibitemShut
  {NoStop}%
\bibitem [{\citenamefont {Aad}\ \emph {et~al.}(2023)\citenamefont {Aad} \emph
  {et~al.}}]{ATLAS:2022qxm}%
  \BibitemOpen
  \bibfield  {author} {\bibinfo {author} {\bibfnamefont {G.}~\bibnamefont
  {Aad}} \emph {et~al.} (\bibinfo {collaboration} {ATLAS}),\ }\href
  {https://doi.org/10.1140/epjc/s10052-023-11699-1} {\bibfield  {journal}
  {\bibinfo  {journal} {Eur. Phys. J.}\ }\textbf {\bibinfo {volume} {C83}},\
  \bibinfo {pages} {681} (\bibinfo {year} {2023})},\ \Eprint
  {https://arxiv.org/abs/2211.16345} {arXiv:2211.16345 [physics.data-an]}
  \BibitemShut {NoStop}%
\bibitem [{\citenamefont {Sirunyan}\ \emph {et~al.}(2018)\citenamefont
  {Sirunyan} \emph {et~al.}}]{CMS:2017wtu}%
  \BibitemOpen
  \bibfield  {author} {\bibinfo {author} {\bibfnamefont {A.~M.}\ \bibnamefont
  {Sirunyan}} \emph {et~al.} (\bibinfo {collaboration} {CMS}),\ }\href
  {https://doi.org/10.1088/1748-0221/13/05/P05011} {\bibfield  {journal}
  {\bibinfo  {journal} {JINST}\ }\textbf {\bibinfo {volume} {13}}\bibfield
  {number} {\bibinfo  {number} { (05)},\ \bibinfo {pages} {P05011}},\ }\Eprint
  {https://arxiv.org/abs/1712.07158} {arXiv:1712.07158 [physics.ins-det]}
  \BibitemShut {NoStop}%
\bibitem [{\citenamefont {Wong}\ \emph {et~al.}(2020)\citenamefont {Wong},
  \citenamefont {Li}, \citenamefont {Brooks}, \citenamefont {Durham},
  \citenamefont {Liu}, \citenamefont {Morreale}, \citenamefont {da~Silva},\
  and\ \citenamefont {Sondheim}}]{Wong:2020xtc}%
  \BibitemOpen
  \bibfield  {author} {\bibinfo {author} {\bibfnamefont {C.-P.}\ \bibnamefont
  {Wong}}, \bibinfo {author} {\bibfnamefont {X.}~\bibnamefont {Li}}, \bibinfo
  {author} {\bibfnamefont {M.}~\bibnamefont {Brooks}}, \bibinfo {author}
  {\bibfnamefont {M.~J.}\ \bibnamefont {Durham}}, \bibinfo {author}
  {\bibfnamefont {M.~X.}\ \bibnamefont {Liu}}, \bibinfo {author} {\bibfnamefont
  {A.}~\bibnamefont {Morreale}}, \bibinfo {author} {\bibfnamefont
  {C.}~\bibnamefont {da~Silva}},\ and\ \bibinfo {author} {\bibfnamefont
  {W.~E.}\ \bibnamefont {Sondheim}},\ }\href@noop {} {\  (\bibinfo {year} {2020})},\
  \Eprint {https://arxiv.org/abs/2009.02888} {arXiv:2009.02888 [nucl-ex]}
  \BibitemShut {NoStop}%
\bibitem [{\citenamefont {Arratia}\ \emph {et~al.}(2021)\citenamefont
  {Arratia}, \citenamefont {Furletova}, \citenamefont {Hobbs}, \citenamefont
  {Olness},\ and\ \citenamefont {Sekula}}]{Arratia:2020azl}%
  \BibitemOpen
  \bibfield  {author} {\bibinfo {author} {\bibfnamefont {M.}~\bibnamefont
  {Arratia}}, \bibinfo {author} {\bibfnamefont {Y.}~\bibnamefont {Furletova}},
  \bibinfo {author} {\bibfnamefont {T.~J.}\ \bibnamefont {Hobbs}}, \bibinfo
  {author} {\bibfnamefont {F.}~\bibnamefont {Olness}},\ and\ \bibinfo {author}
  {\bibfnamefont {S.~J.}\ \bibnamefont {Sekula}},\ }\href
  {https://doi.org/10.1103/PhysRevD.103.074023} {\bibfield  {journal} {\bibinfo
   {journal} {Phys. Rev.}\ }\textbf {\bibinfo {volume} {D103}},\ \bibinfo
  {pages} {074023} (\bibinfo {year} {2021})},\ \Eprint
  {https://arxiv.org/abs/2006.12520} {arXiv:2006.12520 [hep-ph]} \BibitemShut
  {NoStop}%
\bibitem [{\citenamefont {Dong}\ \emph {et~al.}(2023)\citenamefont {Dong},
  \citenamefont {Ji}, \citenamefont {Kelsey}, \citenamefont {Radhakrishnan},
  \citenamefont {Sichtermann},\ and\ \citenamefont {Zhao}}]{Dong:2022xbd}%
  \BibitemOpen
  \bibfield  {author} {\bibinfo {author} {\bibfnamefont {X.}~\bibnamefont
  {Dong}}, \bibinfo {author} {\bibfnamefont {Y.}~\bibnamefont {Ji}}, \bibinfo
  {author} {\bibfnamefont {M.}~\bibnamefont {Kelsey}}, \bibinfo {author}
  {\bibfnamefont {S.}~\bibnamefont {Radhakrishnan}}, \bibinfo {author}
  {\bibfnamefont {E.}~\bibnamefont {Sichtermann}},\ and\ \bibinfo {author}
  {\bibfnamefont {Y.}~\bibnamefont {Zhao}},\ }\href
  {https://doi.org/10.1103/PhysRevD.107.074022} {\bibfield  {journal} {\bibinfo
   {journal} {Phys. Rev.}\ }\textbf {\bibinfo {volume} {D107}},\ \bibinfo
  {pages} {074022} (\bibinfo {year} {2023})},\ \Eprint
  {https://arxiv.org/abs/2210.08609} {arXiv:2210.08609 [hep-ph]} \BibitemShut
  {NoStop}%
\bibitem [{\citenamefont {Aschenauer}\ \emph {et~al.}(2017)\citenamefont
  {Aschenauer}, \citenamefont {Fazio}, \citenamefont {Lamont}, \citenamefont
  {Paukkunen},\ and\ \citenamefont {Zurita}}]{Aschenauer:2017oxs}%
  \BibitemOpen
  \bibfield  {author} {\bibinfo {author} {\bibfnamefont {E.~C.}\ \bibnamefont
  {Aschenauer}}, \bibinfo {author} {\bibfnamefont {S.}~\bibnamefont {Fazio}},
  \bibinfo {author} {\bibfnamefont {M.~A.~C.}\ \bibnamefont {Lamont}}, \bibinfo
  {author} {\bibfnamefont {H.}~\bibnamefont {Paukkunen}},\ and\ \bibinfo
  {author} {\bibfnamefont {P.}~\bibnamefont {Zurita}},\ }
  \href {https://doi.org/10.1103/PhysRevD.96.114005} {\bibfield
  {journal} {\bibinfo  {journal} {Phys. Rev.}\ }\textbf {\bibinfo {volume}
  {D96}},\ \bibinfo {pages} {114005} (\bibinfo {year} {2017})},\ \Eprint
  {https://arxiv.org/abs/1708.05654} {arXiv:1708.05654 [nucl-ex]} \BibitemShut
  {NoStop}%
\bibitem [{\citenamefont {Bierlich}\ \emph {et~al.}(2022)\citenamefont
  {Bierlich} \emph {et~al.}}]{Bierlich:2022pfr}%
  \BibitemOpen
  \bibfield  {author} {\bibinfo {author} {\bibfnamefont {C.}~\bibnamefont
  {Bierlich}} \emph {et~al.},\ }\href
  {https://doi.org/10.21468/SciPostPhysCodeb.8} {\bibfield  {journal} {\bibinfo
   {journal} {SciPost Phys. Codeb.}\ }\textbf {\bibinfo {volume} {2022}},\
  \bibinfo {pages} {8} (\bibinfo {year} {2022})},\ \Eprint
  {https://arxiv.org/abs/2203.11601} {arXiv:2203.11601 [hep-ph]} \BibitemShut
  {NoStop}%
\bibitem [{Note1()}]{Note1}%
  \BibitemOpen
  \bibinfo {note} {Natural units are used throughout this paper.}\BibitemShut
  {Stop}%
\bibitem [{\citenamefont {Armesto}\ \emph {et~al.}(2023)\citenamefont
  {Armesto}, \citenamefont {Cridge}, \citenamefont {Giuli}, \citenamefont
  {Harland-Lang}, \citenamefont {Newman}, \citenamefont {Schmookler},
  \citenamefont {Thorne},\ and\ \citenamefont {Wichmann}}]{Armesto:2023hnw}%
  \BibitemOpen
  \bibfield  {author} {\bibinfo {author} {\bibfnamefont {N.}~\bibnamefont
  {Armesto}}, \bibinfo {author} {\bibfnamefont {T.}~\bibnamefont {Cridge}},
  \bibinfo {author} {\bibfnamefont {F.}~\bibnamefont {Giuli}}, \bibinfo
  {author} {\bibfnamefont {L.}~\bibnamefont {Harland-Lang}}, \bibinfo {author}
  {\bibfnamefont {P.}~\bibnamefont {Newman}}, \bibinfo {author} {\bibfnamefont
  {B.}~\bibnamefont {Schmookler}}, \bibinfo {author} {\bibfnamefont
  {R.}~\bibnamefont {Thorne}},\ and\ \bibinfo {author} {\bibfnamefont
  {K.}~\bibnamefont {Wichmann}},\ }\href@noop {} {\  (\bibinfo {year} {2023})},\ \Eprint
  {https://arxiv.org/abs/2309.11269} {arXiv:2309.11269 [hep-ph]} \BibitemShut
  {NoStop}%
\bibitem [{\citenamefont {Williams}\ \emph {et~al.}(2011)\citenamefont
  {Williams}, \citenamefont {Gligorov}, \citenamefont {Thomas}, \citenamefont
  {Dijkstra}, \citenamefont {Nardulli},\ and\ \citenamefont
  {Spradlin}}]{Williams:2011aza}%
  \BibitemOpen
  \bibfield  {author} {\bibinfo {author} {\bibfnamefont {M.}~\bibnamefont
  {Williams}}, \bibinfo {author} {\bibfnamefont {V.}~\bibnamefont {Gligorov}},
  \bibinfo {author} {\bibfnamefont {C.}~\bibnamefont {Thomas}}, \bibinfo
  {author} {\bibfnamefont {H.}~\bibnamefont {Dijkstra}}, \bibinfo {author}
  {\bibfnamefont {J.}~\bibnamefont {Nardulli}},\ and\ \bibinfo {author}
  {\bibfnamefont {P.}~\bibnamefont {Spradlin}},\ }\href@noop {} {\  (\bibinfo {year}
  {2011})}\BibitemShut {NoStop}%
\bibitem [{\citenamefont {Abe}\ \emph {et~al.}(1998)\citenamefont {Abe} \emph
  {et~al.}}]{SLD:1997ihw}%
  \BibitemOpen
  \bibfield  {author} {\bibinfo {author} {\bibfnamefont {K.}~\bibnamefont
  {Abe}} \emph {et~al.} (\bibinfo {collaboration} {SLD}),\ }\href
  {https://doi.org/10.1103/PhysRevLett.80.660} {\bibfield  {journal} {\bibinfo
  {journal} {Phys. Rev. Lett.}\ }\textbf {\bibinfo {volume} {80}},\ \bibinfo
  {pages} {660} (\bibinfo {year} {1998})},\ \Eprint
  {https://arxiv.org/abs/hep-ex/9708015} {arXiv:hep-ex/9708015} \BibitemShut
  {NoStop}%
\bibitem [{\citenamefont {Aaij}\ \emph
  {et~al.}(2015{\natexlab{b}})\citenamefont {Aaij} \emph
  {et~al.}}]{LHCb:2015bwt}%
  \BibitemOpen
  \bibfield  {author} {\bibinfo {author} {\bibfnamefont {R.}~\bibnamefont
  {Aaij}} \emph {et~al.} (\bibinfo {collaboration} {LHCb}),\ }
  \href {https://doi.org/10.1103/PhysRevD.92.052001}
  {\bibfield  {journal} {\bibinfo  {journal} {Phys. Rev.}\ }\textbf {\bibinfo
  {volume} {D92}},\ \bibinfo {pages} {052001} (\bibinfo {year}
  {2015}{\natexlab{b}})},\ \Eprint {https://arxiv.org/abs/1505.04051}
  {arXiv:1505.04051 [hep-ex]} \BibitemShut {NoStop}%
\bibitem [{\citenamefont {Aaij}\ \emph
  {et~al.}(2015{\natexlab{c}})\citenamefont {Aaij} \emph
  {et~al.}}]{LHCb:2015nta}%
  \BibitemOpen
  \bibfield  {author} {\bibinfo {author} {\bibfnamefont {R.}~\bibnamefont
  {Aaij}} \emph {et~al.} (\bibinfo {collaboration} {LHCb}),\ }
  \href {https://doi.org/10.1103/PhysRevLett.115.112001}
  {\bibfield  {journal} {\bibinfo  {journal} {Phys. Rev. Lett.}\ }\textbf
  {\bibinfo {volume} {115}},\ \bibinfo {pages} {112001} (\bibinfo {year}
  {2015}{\natexlab{c}})},\ \Eprint {https://arxiv.org/abs/1506.00903}
  {arXiv:1506.00903 [hep-ex]} \BibitemShut {NoStop}%
\bibitem [{\citenamefont {Candido}\ \emph {et~al.}(2024)\citenamefont
  {Candido}, \citenamefont {Hekhorn}, \citenamefont {Magni}, \citenamefont
  {Rabemananjara},\ and\ \citenamefont {Stegeman}}]{Candido:2024rkr}%
  \BibitemOpen
  \bibfield  {author} {\bibinfo {author} {\bibfnamefont {A.}~\bibnamefont
  {Candido}}, \bibinfo {author} {\bibfnamefont {F.}~\bibnamefont {Hekhorn}},
  \bibinfo {author} {\bibfnamefont {G.}~\bibnamefont {Magni}}, \bibinfo
  {author} {\bibfnamefont {T.~R.}\ \bibnamefont {Rabemananjara}},\ and\
  \bibinfo {author} {\bibfnamefont {R.}~\bibnamefont {Stegeman}},\ }
  \href@noop {} {\  (\bibinfo {year} {2024})},\ \Eprint
  {https://arxiv.org/abs/2401.15187} {arXiv:2401.15187 [hep-ph]} \BibitemShut
  {NoStop}%
\bibitem [{\citenamefont {Buckley}\ \emph {et~al.}(2015)\citenamefont
  {Buckley}, \citenamefont {Ferrando}, \citenamefont {Lloyd}, \citenamefont
  {Nordstr\"om}, \citenamefont {Page}, \citenamefont {R\"ufenacht},
  \citenamefont {Sch\"onherr},\ and\ \citenamefont {Watt}}]{Buckley:2014ana}%
  \BibitemOpen
  \bibfield  {author} {\bibinfo {author} {\bibfnamefont {A.}~\bibnamefont
  {Buckley}}, \bibinfo {author} {\bibfnamefont {J.}~\bibnamefont {Ferrando}},
  \bibinfo {author} {\bibfnamefont {S.}~\bibnamefont {Lloyd}}, \bibinfo
  {author} {\bibfnamefont {K.}~\bibnamefont {Nordstr\"om}}, \bibinfo {author}
  {\bibfnamefont {B.}~\bibnamefont {Page}}, \bibinfo {author} {\bibfnamefont
  {M.}~\bibnamefont {R\"ufenacht}}, \bibinfo {author} {\bibfnamefont
  {M.}~\bibnamefont {Sch\"onherr}},\ and\ \bibinfo {author} {\bibfnamefont
  {G.}~\bibnamefont {Watt}},\ }\href
  {https://doi.org/10.1140/epjc/s10052-015-3318-8} {\bibfield  {journal}
  {\bibinfo  {journal} {Eur. Phys. J.}\ }\textbf {\bibinfo {volume} {C75}},\
  \bibinfo {pages} {132} (\bibinfo {year} {2015})},\ \Eprint
  {https://arxiv.org/abs/1412.7420} {arXiv:1412.7420 [hep-ph]} \BibitemShut
  {NoStop}%
\bibitem [{\citenamefont {Cerci}\ \emph {et~al.}(2023)\citenamefont {Cerci},
  \citenamefont {Demiroglu}, \citenamefont {Deshpande}, \citenamefont {Newman},
  \citenamefont {Schmookler}, \citenamefont {Sunar~Cerci},\ and\ \citenamefont
  {Wichmann}}]{Cerci:2023uhu}%
  \BibitemOpen
  \bibfield  {author} {\bibinfo {author} {\bibfnamefont {S.}~\bibnamefont
  {Cerci}}, \bibinfo {author} {\bibfnamefont {Z.~S.}\ \bibnamefont
  {Demiroglu}}, \bibinfo {author} {\bibfnamefont {A.}~\bibnamefont
  {Deshpande}}, \bibinfo {author} {\bibfnamefont {P.~R.}\ \bibnamefont
  {Newman}}, \bibinfo {author} {\bibfnamefont {B.}~\bibnamefont {Schmookler}},
  \bibinfo {author} {\bibfnamefont {D.}~\bibnamefont {Sunar~Cerci}},\ and\
  \bibinfo {author} {\bibfnamefont {K.}~\bibnamefont {Wichmann}},\ }\href
  {https://doi.org/10.1140/epjc/s10052-023-12176-5} {\bibfield  {journal}
  {\bibinfo  {journal} {Eur. Phys. J.}\ }\textbf {\bibinfo {volume} {C83}},\
  \bibinfo {pages} {1011} (\bibinfo {year} {2023})},\ \Eprint
  {https://arxiv.org/abs/2307.01183} {arXiv:2307.01183 [hep-ph]} \BibitemShut
  {NoStop}%
\bibitem [{\citenamefont {Cridge}\ \emph {et~al.}(2021)\citenamefont {Cridge},
  \citenamefont {Harland-Lang}, \citenamefont {Martin},\ and\ \citenamefont
  {Thorne}}]{Cridge:2021qfd}%
  \BibitemOpen
  \bibfield  {author} {\bibinfo {author} {\bibfnamefont {T.}~\bibnamefont
  {Cridge}}, \bibinfo {author} {\bibfnamefont {L.~A.}\ \bibnamefont
  {Harland-Lang}}, \bibinfo {author} {\bibfnamefont {A.~D.}\ \bibnamefont
  {Martin}},\ and\ \bibinfo {author} {\bibfnamefont {R.~S.}\ \bibnamefont
  {Thorne}},\ }\href {https://doi.org/10.1140/epjc/s10052-021-09533-7}
  {\bibfield  {journal} {\bibinfo  {journal} {Eur. Phys. J.}\ }\textbf
  {\bibinfo {volume} {C81}},\ \bibinfo {pages} {744} (\bibinfo {year} {2021})},\
  \Eprint {https://arxiv.org/abs/2106.10289} {arXiv:2106.10289 [hep-ph]}
  \BibitemShut {NoStop}%
\bibitem [{\citenamefont {Campbell}\ \emph {et~al.}(2021)\citenamefont
  {Campbell}, \citenamefont {Neumann},\ and\ \citenamefont
  {Sullivan}}]{Campbell:2021qgd}%
  \BibitemOpen
  \bibfield  {author} {\bibinfo {author} {\bibfnamefont {J.}~\bibnamefont
  {Campbell}}, \bibinfo {author} {\bibfnamefont {T.}~\bibnamefont {Neumann}},\
  and\ \bibinfo {author} {\bibfnamefont {Z.}~\bibnamefont {Sullivan}},\
  }\href
  {https://doi.org/10.1103/PhysRevD.104.094042} {\bibfield  {journal} {\bibinfo
   {journal} {Phys. Rev.}\ }\textbf {\bibinfo {volume} {D104}},\ \bibinfo
  {pages} {094042} (\bibinfo {year} {2021})},\ \Eprint
  {https://arxiv.org/abs/2109.10448} {arXiv:2109.10448 [hep-ph]} \BibitemShut
  {NoStop}%
\bibitem [{\citenamefont {Accardi}\ \emph {et~al.}(2016)\citenamefont {Accardi}
  \emph {et~al.}}]{Accardi:2016ndt}%
  \BibitemOpen
  \bibfield  {author} {\bibinfo {author} {\bibfnamefont {A.}~\bibnamefont
  {Accardi}} \emph {et~al.},\ }\href
  {https://doi.org/10.1140/epjc/s10052-016-4285-4} {\bibfield  {journal}
  {\bibinfo  {journal} {Eur. Phys. J.}\ }\textbf {\bibinfo {volume} {C76}},\
  \bibinfo {pages} {471} (\bibinfo {year} {2016})},\ \Eprint
  {https://arxiv.org/abs/1603.08906} {arXiv:1603.08906 [hep-ph]} \BibitemShut
  {NoStop}%
\bibitem [{\citenamefont {Li}\ \emph {et~al.}(2022)\citenamefont {Li},
  \citenamefont {Liu},\ and\ \citenamefont {Vitev}}]{Li:2021gjw}%
  \BibitemOpen
  \bibfield  {author} {\bibinfo {author} {\bibfnamefont {H.~T.}\ \bibnamefont
  {Li}}, \bibinfo {author} {\bibfnamefont {Z.~L.}\ \bibnamefont {Liu}},\ and\
  \bibinfo {author} {\bibfnamefont {I.}~\bibnamefont {Vitev}},\ }\href
  {https://doi.org/10.1016/j.physletb.2022.137007} {\bibfield  {journal}
  {\bibinfo  {journal} {Phys. Lett.}\ }\textbf {\bibinfo {volume} {B827}},\
  \bibinfo {pages} {137007} (\bibinfo {year} {2022})},\ \Eprint
  {https://arxiv.org/abs/2108.07809} {arXiv:2108.07809 [hep-ph]} \BibitemShut
  {NoStop}%
\bibitem [{\citenamefont {del Castillo}\ \emph {et~al.}(2022)\citenamefont {del
  Castillo}, \citenamefont {Echevarria}, \citenamefont {Makris},\ and\
  \citenamefont {Scimemi}}]{delCastillo:2021znl}%
  \BibitemOpen
  \bibfield  {author} {\bibinfo {author} {\bibfnamefont {R.~F.}\ \bibnamefont
  {del Castillo}}, \bibinfo {author} {\bibfnamefont {M.~G.}\ \bibnamefont
  {Echevarria}}, \bibinfo {author} {\bibfnamefont {Y.}~\bibnamefont {Makris}},\
  and\ \bibinfo {author} {\bibfnamefont {I.}~\bibnamefont {Scimemi}},\
  }\href
  {https://doi.org/10.1007/JHEP03(2022)047} {\bibfield  {journal} {\bibinfo
  {journal} {JHEP}\ }\textbf {\bibinfo {volume} {03}},\ \bibinfo {pages}
  {047}},\ \Eprint {https://arxiv.org/abs/2111.03703} {arXiv:2111.03703
  [hep-ph]} \BibitemShut {NoStop}%
\end{thebibliography}%

\end{document}